\documentclass{moriond}

\usepackage{amsmath,amssymb}
\usepackage{graphicx}

\def\mco{\multicolumn}

\def\be{\begin{equation}}
\def\ee{\end{equation}}
\def\bea{\begin{eqnarray}}
\def\eea{\end{eqnarray}}

\usepackage[normalem]{ulem} 
\newcommand{\Photo}{}

\begin{document}
\vspace*{4cm}
\title{Search for cosmic rays in GRANDProto300}

\author{Jolan Lavoisier, Aurélien Benoit-Lévy, Kumiko Kotera, Mauricio Bustamante, Marion Guelfand, Nathan Lebas, Takashi Sako and Xishui Tian, on behalf of the GRAND collaboration.}

\address{Institut d'Astrophysique de Paris, CNRS UMR 7095, Sorbonne Université,\\ 98 bis Boulevard Arago, 75014 Paris, France,\\
ILANCE, CNRS - University of Tokyo International Research Laboratory,\\
Kashiwa, Chiba 277-8582, Japan}

\maketitle\abstracts{
GRANDProto300 is a prototype array of the GRAND experiment, designed to validate the technique of autonomous radio-detection of astroparticles by detecting cosmic rays with energies between 10$^{17} - 10^{18.5}$ eV. This observation will further enable the study of the Galactic-to-extragalactic source transition region. Between November 2024 and June 2025, 46 antennas were operational, and since July 2025, 63 antennas have been collecting data stably. We present here our cosmic-ray search pipeline, that involves several filtering steps: (1) exclusion of clustered (in time and space) noise events, (2) polarization cut, (3) zenith arrival cut, and (4) delay in expected trigger cut. The efficiency is evaluated on the latest batch of data from July 2025 to October 2025. Additionally, we present an identification layer that is being developed: a CNN-based algorithm to discriminate between noise and cosmic-ray signals, using signal traces. We present the accuracy of our model, trained both on simulations and on real cosmic-ray signal datasets.}

\section{Introduction}
\subsection{The Giant Radio Array for Neutrino Detection}
The Giant Radio Array for Neutrino Detection (GRAND) experiment aims to detect cosmic particles with energies exceeding $10^{17}$ eV with an array of radio antennas~\cite{omaICRC2025}. GRANDProto300 (GP300) is a prototype array of the GRAND experiment, that consists of 300 antennas spread over an area of 200\,km$^2$ situated in the Gobi desert, near Dunhuang, China (located at $40.99434^\circ$N, $93.94177^\circ$E), designed to validate the technique of autonomous radio-detection of astroparticles.
GP300 is specifically aimed at detecting cosmic rays with energies between $10^{17}-10^{18.5}$\,eV. This energy range is crucial for studying the transition region between Galactic and extragalactic sources of cosmic rays.

Within the GRAND collaboration and in the GRANDlib software framework~\cite{GRANlib24}, a data analysis pipeline is being developed to identify cosmic rays and analyze their properties, among the dataset of the first stable GP300 prototype runs. 

\subsection{Experimental setup}\label{Moriond:setup}
 As of July 2025, 65 antennas with an infill step of 577 m have been deployed and 63 are operational, collecting data stably \cite{GRANDprotoHW_25}. Each Detection Unit (DU) operates with a sampling frequency of 500\,MHz and an analog-to-digital (ADC) converter digitizes signals across three distinct channels, each constructed from a dipole arm of the antenna, along a direction ($X$ channel along North-South, $Y$ channel along West-East, and $Z$ channel along the vertical axis). The ADC data, recorded over a duration of 2\,$\mu$s in a frequency range of 50 to 200\,MHz, is then converted to voltage units. The GRAND experiment handles data in three formats: Monitoring Data (MD), continuous monitoring of the background radio environments to identify and filter out noise; Unit Data (UD), which records signals from individual DUs that pass a local trigger at DU level based on 6 parameters, including two amplitude thresholds set at 5 and 4 sigma above background noise, ensuring that only transient radio pulses are captured; Coincidence Data (CD), data from multiple ($\geq$5) antennas triggered within a 13\,$\mu$s time window, providing a more accurate identification and reconstruction of the signal. The collected data is processed via the GRANDlib software \cite{GRANlib24} in an offline treatment.

The data periods used for the search of cosmic rays presented in this proceeding cover July through October 2025. The trigger parameters remained unchanged during the period, but the layout experimented bursts in data during July and August, and 75\% of total events were recorded during this period. In total, our dataset comprises 5,477,117 events ({\it i.e.} CD event).

\section{Cosmic-ray search pipeline}
The events are subject to two types of noise: ``parasites" --transient noise inducing triggering of a false event--, and stationary background noise --appearing within the baseline of the trace. Parasites are targeted for systematic removal due to their high occurrence. The cosmic search pipeline aims at identifying cosmic rays in the batch of events by selecting the data through quality and physical cuts. For this pipeline, we adopt a conservative approach and prioritize purity over efficiency in our identification process.

\subsection{Systematic cuts}
\textbf{Cluster cut}: The GP300 dataset is affected by interference from commercial aircraft and emissions from an electric transformer northeast of the site, resulting in bursts of parasites that can be excluded from the analysis. Given that cosmic rays produce short radio pulses ($\lesssim 100$\,ns) and thus cannot result in multiple events, a clustering algorithm is well-suited for this task. The algorithm is based on the spatial proximity of events using Planar Wave Front (PWF) arrival direction reconstruction~\cite{Ferri_re_2025}. Multiple events occurring within the same angular and time window are designated as a cluster and consequently excluded from the dataset. The angular distance $\Delta \vartheta$ between events $A$ and $B$ is defined as the minimum between $\Delta\theta=|\theta_B - \theta_A|$ and $\Delta\varphi =|([\varphi_B-\varphi_A+\pi]\,\mod\,2\pi) -\pi|$ and their temporal distance as $\Delta t= |t_B-t_A|$. By varying the angular and time cut parameters $(\Delta \vartheta_{\rm cut}, \Delta t_{\rm cut})$ below which the events are suppressed, we adopt a conservative approach and retain the cut parameters of $\Delta \vartheta_{\rm cut} = \min(\Delta \theta, \Delta \phi)=5^\circ$ and $\Delta t_{\rm cut}=5\,$. This results in a cut of 88\,\% of the CD data.

\textbf{Polarization cut}: When a cosmic-ray air shower is created in the atmosphere, the geomagnetic field ${\bf B}$, through the Lorentz force, will act on the charged secondary particles and create a linear electric field along $\bf{k}\times\bf{B}$ where $\bf{k}$ is the particle velocity: this specific polarization signature along $\bf{k}\times\bf{B}$ can be used to identify cosmic-ray signals. In particular~\cite{Chiche_2022}, we can use the component of the electric field along the {\bf B} direction ${\bf e}_V\cdot{\bf e}_B$ as an estimator of polarization, where $\bf{e}_V$ is the normalized voltage vector from the antenna, $\bf{e}_B$ is the normalized geomagnetic field vector. ${\bf e}_V\cdot{\bf e}_B =0$ indicates that the electric field is orthogonal to $\bf{B}$, which is the case for cosmic rays. This method demonstrates powerful identification efficiency: for more details about the method, see~\cite{jlavICRC2025}. For our current data batch, this cuts 35\% of total events.

\textbf{Zenith cut}: The antenna layout of GP300 renders the detector efficient in highly-inclined-air-shower detection. Using PWF reconstruction, we remove events with a reconstructed arrival direction $\theta\le 60^\circ$ in zenith angle. In addition, reconstruction of near-horizon event is difficult, hence for now we remove as well events with zenith angle $\theta\ge 88^\circ$. This cut labels as parasites 61\% of the full event batch.

\textbf{Delay cut}: The Spherical Wave Fit (SWF) reconstruction method enables precise localization of the emission source for a given event. By determining the source point, we can systematically compare the recorded antenna detection times with the theoretically expected arrival times for an electromagnetic signal originating from this location. The discrepancy between measured and expected times is quantified using a $\chi^2$ metric, defined as $\chi^2 = 1/N_{\rm ant}\sum (t_{\rm mes} - t_{\rm exp})^2/\sigma^2$, where $N_{\rm ant}$ is the number of antennas in the event, $t_{\rm mes}$ the measured time at each antenna, $t_{\rm exp}$ the expected time at each antenna for an EM emission and $\sigma = 5$\,ns the time resolution of antennas. Implementing a $\chi^2$ threshold of 10$^3$ filters out 44\% of the data.

\textbf{Additional quality cuts}: \label{Moriond:add_cuts}
The previous cuts enable to filter out 99.63\% of the parasites in the event batch. The remaining $\sim 20\,000$ candidates are subsequently analyzed and tentatively reconstructed using the Angular Distribution Fit (ADF)~\cite{ADF}~\cite{mguelMoriond2026}. This method outputs a $\chi^2_{\rm ADF}$ value as an indicator of the quality of reconstruction, on which we place a maximum illustrative threshold of 20. 

The current selection process does not account for
the shape of the recorded traces (electric voltage as a function of time): cosmic-ray traces should display specific short pulses at low frequencies. While a visual selection criterion has been temporarily applied, efforts are underway to develop systematic cuts, as discussed in Section~\ref{Moriond:ML_work}.

\subsection{Cosmic-ray candidates}
\begin{figure}
    \centering
    \noindent\begin{minipage}{0.4\textwidth}
    \includegraphics[width=\textwidth, trim={2.3cm 0 .5cm 0}, clip]{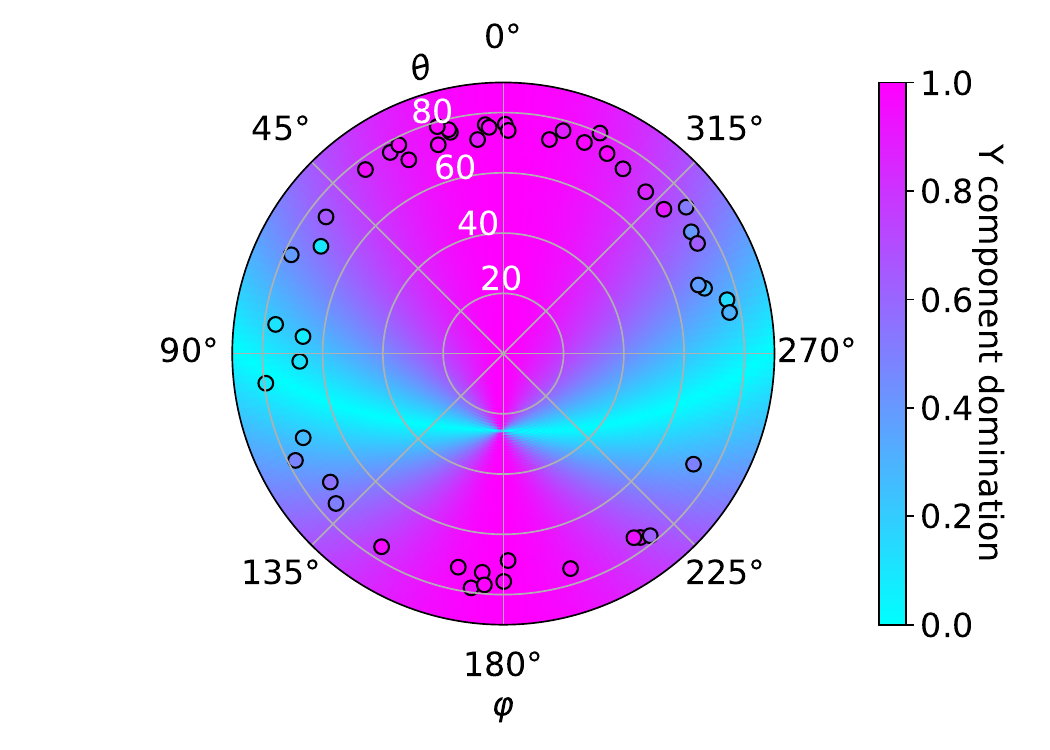}
    \end{minipage}
    \noindent\begin{minipage}{0.55\textwidth}
    \resizebox{\textwidth}{!}{%
        \begin{tabular}{|l|ccc|ccc|}\hline
            & \mco{3}{|c|}{Model 1D} & \mco{3}{|c|}{Model 2D} \\ \hline
            & Parasites & Sims & Real CR & Parasites & Sims & Real CR \\ \hline
            Pred $\geq$ 0.5 & 117 & 1206 & 226 & 190 & 1233 & 259\\
            Pred $\leq$ 0.5 & 1132 & 54 & 138 & 1059 & 27 & 105\\
            Recall & & 95.71\% & 62.08\% & & 97.85\% & 71.15\% \\ \hline
        \end{tabular}}
        \caption{\footnotesize \textbf{Left:} Arrival directions (zenith $\theta$ and azimuth $\phi$ angles where $\theta=0$ indicates up and $\phi=0$ indicates north) of the cosmic-ray candidates in polar representation. The colormap indicates the ratio $\max Y/ [(\max X)^2+(\max Y)^2+(\max Z)^2]^{1/2}$ for candidate events and, in the background, expected for incoming cosmic rays. \textbf{Right}: Prediction of the classification models on test sets.}
        \label{Moriond:CRC_MLresults}
    \end{minipage}
    \vspace{-0.3cm}
\end{figure}

The application of the previously described selection criteria results in a selection of 51 cosmic-ray candidates. Fig.~\ref{Moriond:CRC_MLresults} left shows their arrival direction as well as the dominance of the signal in the East-West direction, as theoretically expected given the geomagnetic field configuration (see caption). The electromagnetic energy of these candidates was reconstructed and presented in this proceeding series~\cite{mguelMoriond2026}~\cite{aferrMoriond2026}. 

\section{Machine Learning model for pulse discrimination}\label{Moriond:ML_work}
The NUTRIG project~\cite{pcorrICRC2025} enables to take into account the pulse shape in the recorded traces by template fitting, for more efficient online triggering. We explore here an alternative offline technique, using a machine learning classification model, following pioneering works~\cite{slecozICRC2023}~\cite{IceCube}.

\subsection{Dataset building}

To train our binary classification model, we construct two datasets: a parasite dataset and a cosmic-ray dataset (composed of simulation pulses and real cosmic-ray pulses), both derived from the same experimental data batches recorded in July 2025\footnote{excluding the aforementioned cosmic rays. It is assumed here that the remaining batch contains only parasites.}. This batch is divided into three equal parts.

The first part is used to create the parasite dataset by randomly selecting 384-point traces containing parasitic pulses. The second part serves as background noise for the simulation dataset: 384-point noise traces are extracted, and a simulated cosmic-ray pulse is inserted at a random position. These traces are then processed through the experimental trigger (see Section \ref{Moriond:setup}). The third part is reserved for generating test datasets for both parasitic and simulated signals. Furthermore, to prevent simulation gap, we complete the cosmic-ray dataset with the real cosmic rays identified earlier, applying the same extraction method as for the parasites, obtaining 567 traces. To have a sufficient amount of training data, we augment the data by randomizing multiple times the pulse position in its trace.

All extracted traces are required to have a background noise level between 10 and 20 ADC and a signal-to-noise ratio (SNR) exceeding 4. The input traces consist of 384 points across the two channels X (North-South) and Y (East-West), as cosmic-ray signals are predominantly recorded in these directions.

\subsection{Model}
The preliminary architecture for this analysis consists of 13 convolutional layers with varying filter counts (64, four layers of 16, 32, and 64), each followed by batch normalization and ReLU activation, and concluding with global average pooling and a single-neuron dense layer. Given the two-channel input (X and Y), we evaluate both one-dimensional and two-dimensional convolutional layers, resulting in two distinct models based on the same architecture.

The loss function adapted to our study is the binary loss function: $\mathfrak{L} = - 1/k\sum_{i=1}^{k}[p_1\log(q_1) + (1-p_1)\log(1-q_1)]_i$, with $k$ the number of data points, $p_1$ the true probability (i.e. true classification) and $q_1$ estimated probability (predicted probability of observation). We obtain our results by training on 4000 epochs, with a batch size of 128 and a learning rate of 0.01.

\subsection{Results}
Both models are trained using 1448 cosmic-ray traces, 1000 simulated traces and 2497 parasites. Testing is performed on 364 cosmic-ray traces, 1260 simulated traces and 1249 parasites with results summarized in the top table of Fig.~\ref{Moriond:CRC_MLresults}. To prioritize high cosmic-ray identification efficiency, we focus on the recall metric, defined as $={\rm TP/ (TP + FN)}$.  For simulated traces, recall reaches 95.7\% and 97.9\%, while for real cosmic-ray traces, it is 62.1\% and 71.2\%.

\footnotesize{

\section*{Acknowledgments}

This work was supported by the CNRS Programme Blanc MITI (``GRAND'' 2023.1 268448), the NuTRIG project, Agence Nationale de la Recherche (ANR-21-CE31-0025), the HERON project, Agence Nationale de la Recherche (ANR-24-MRS2-0014). JL acknowledges support from the PhD Joint program as part of the IRC between CNRS-U Tokyo. JL and KK thank ICRR and ILANCE for the hospitality and the support. KK acknowledges support from the Fulbright-France program.
}

\section*{References}
\footnotesize
\bibliography{lavoisier}

\end{document}